\documentclass[conference]{IEEEtran}
\IEEEoverridecommandlockouts
\usepackage{cite}
\usepackage{amsmath,amssymb,amsfonts}
\usepackage{algorithmic}
\usepackage{graphicx}
\usepackage{textcomp}
\usepackage{xcolor}
\def\BibTeX{{\rm B\kern-.05em{\sc i\kern-.025em b}\kern-.08em
    T\kern-.1667em\lower.7ex\hbox{E}\kern-.125emX}}
\begin{document}

\title{MetaAnalysis of Methods for Scaling Blockchain Technology for Automotive Uses
}

\author{\IEEEauthorblockN{Sid Masih}
\IEEEauthorblockA{\textit{Mobility Open Blockchain Initiative (MOBI)} \\
Berkeley, USA \\
sid.masih@berkeley.edu}
\and
\IEEEauthorblockN{ Parth Singhal}
\IEEEauthorblockA{\textit{Mobility Open Blockchain Initiative (MOBI)} \\
Berkeley, USA \\
parthsinghal@berkeley.edu}
}

\maketitle

\begin{abstract}
The automotive industry has seen an increased need for connectivity, both as a result of the advent of autonomous driving and the rise of connected cars and truck fleets. This shift has led to  issues such as trusted coordination and a wider attack surface have come to light, leading to higher costs and bureaucratic intervention . Due to the increasing adoption of connected vehicles, as well as other connected infrastructure, trustless peer to peer systems including blockchain are being explored as potential solutions to this efficiency problem. All the while, scalability is still a significant concern for industry players. Current blockchain based systems have difficulty scaling: Bitcoin can only process seven transactions per second (tx/s) whereas Ethereum’s fifteen tx/s is not a major improvement. Combined with the high cost of consensus and low throughput, such platforms are unusable in the mobility sector. This paper will address the latest advances in the field that aim to resolve parts of this problem as well as inform its readers about the latest scalability technologies that could push blockchain automotive infrastructure into the mainstream. This paper will also introduce the theoretical tools and advancements that, if implemented, could bring the mobility industry closer toward adopting efficient, scalable, and cost effective decentralized solutions.
\end{abstract}

\section{Introduction}
Recent advancements in automotive communication technologies have led connected vehicles and their infrastructures to the internet as a solution. Many automakers have adopted cellular modems into vehicles, enabling a rich set of infrastructure connectivity services such as ad-hoc vehicle to everything (V2X), VANETs \cite{b1} \cite{b2}, LTE-Advanced \cite{b3},  and high throughput Millimeter Wave \cite{b63} connections. This has led to a vast number of Internet of Things (IoT) like applications for vehicles and infrastructure including payments, enhanced infotainment, fleet management, and other solutions, which require infrastructure or vehicle connectivity. Consequently, these applications have resulted in a large set of proposed and forward thinking infrastructure applications including identity systems, peer-to peer payment systems, autonomous fleet control, and other centralized and decentralized infrastructure. However, with each manufacturer attempting to create different competing technologies, and with a lack of widespread standards, the current system is uncoordinated, non-compatible, and wasteful.

Decentralized systems have been proposed as an alternative to share and regulate vehicular infrastructure and streamline complex trusted single entity systems. However, current decentralized blockchain based systems such as Bitcoin and Ethereum are unscalable \cite{b4}. . This paper performs a meta-analysis of recent scalability work that is applicable to potential mobility-focused blockchain architecture. CAP Theorem\cite{b5} \cite{b6} suggests that a distributed system can only have two of the three: Consistency, Availability and Partition Tolerance. Together they form pillars of architecture for a scalable, robust system. Our core assumptions encourage locality of validation (consistency), toleration of network failures (partial tolerance), as well as high transaction throughput (availability). This paper explores the following scalability areas: network enhancements, sharding mechanisms, relevant alternative advancements in consensus protocols, and state channels.


\section{Description of Automotive Environment and Problems}

	             \begin{figure}
            \begin{center}
                 \includegraphics[width=3.3in,height=3.6in,clip,keepaspectratio]{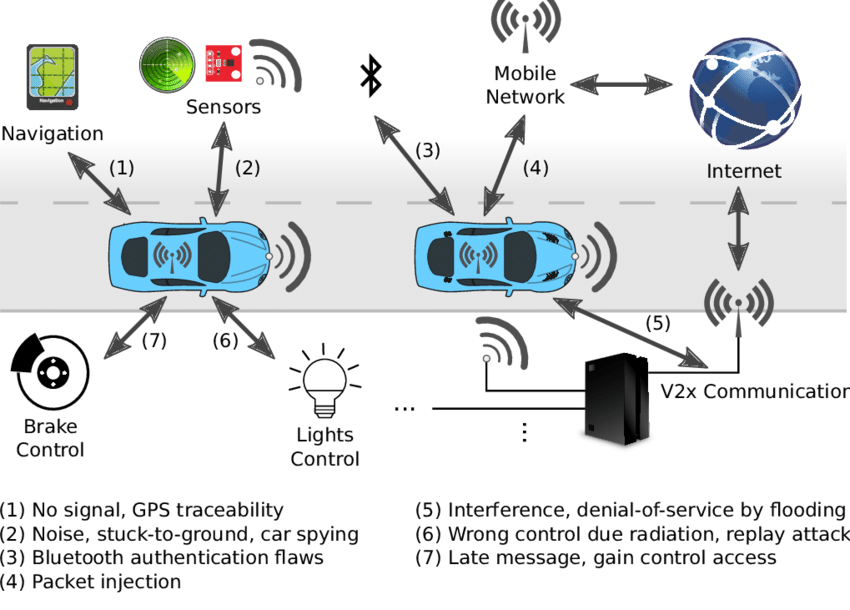}
             \end{center}
            \caption{Connected vehicle environment as illustrated in [62, Fig. 1]}
        \end{figure}

\subsection{Vehicular Environment}

Vehicles have the ability to carry larger and more powerful hardware compared to traditional mobile devices such as cellphones\cite{b7} \cite{b8} \cite{b9}. Furthermore, automakers and tier-1 suppliers are able to carefully control vehicle hardware as well as directly limit access and interfaces to the hardware and other base infrastructure. This creates a reduced attack surface to many potential attacks that require privileged access. Furthermore, it is now common for vehicles to have internet and vehicle to vehicle (V2V) connectivity \cite{b10} \cite{b11}. This critical fact makes decentralized systems now a possibility. 
\subsection{Current Problems}
While much general scalability research has been done, automotive network conditions present unique challenges. Factors such as network connectivity are often ignored, not addressing lossy V2X connections as well as participant mobility often times change how a network behaves \cite{b12}. Additionally, liveness and synchrony \cite{b13} pose difficult requirements for temporal connections, where connectivity bounds are either expensive \cite{b61} or can face delays and message drops. Furthermore connections often degrade at high speeds due to limitations with radio frequency (RF) engineering. Importantly by definition, each vehicle should be treated as mobile. For automotive use cases, geographical changes can become a major bottleneck for non-mobile optimized networks. As such validator discovery and effective load balancing become significant issues. Finally, many blockchain systems are unable to scale with Bitcoin and Ethereum's low transaction throughput compared to Visa's amortized 1700 tx/s processing speed\cite{b4}. With approximately one billion vehicles on roads as of 2019, current systems will not be able to meet throughput requirements. 

\subsection{Liveness}
Liveness \cite{b13} is defined as the ability for a transaction to eventually be  accepted by all nodes in the system. This property is especially challenging for mobility and automotive architectures due to the physical mobility of vehicles and potentially inconsistent network conditions described in Section II, Part A. Additionally, the liveness of a system determines its usability with its actual or amortized tx/s. This determines the actual throughput of system. As such, validator placement and connectivity play crucial roles with ensuring system liveness. Below, Sections IV and VI (Network Enhancements and Consensus Models) play crucial roles in liveness. 

\section{Network enhancements}
\subsection{Overview}
The idea of reducing network latency between two endpoints is an established research direction in both traditional distributed systems and decentralized systems. Such an approach for decentralized ledger systems in mobile environments can decrease orphan blocks  \cite{b27} \cite{b28}, two valid but competing blocks appending to the same network. Since one simple method of scaling is increasing the size of a block, therefore directly increasing the number of transactions that may be added to a block, this will result in increased time for such a block to propagate to the entire network. Reducing the amount of time between blocks being mined (inter-blocktime) results in a higher chance of a fork, as not enough validator nodes receive the new block to append,  discards other incorrect blocks. Multiple approaches have been developed to propagate larger and more efficient blocks at a lower latency to all validators, thus decreasing the block propagation time. In essence, network enhancements allow larger blocks to propagate faster, thus safely increasing effective transaction throughput.

\subsection{Relay Networks}
In blockchain, relay networks are a type of content delivery network (CDN)\cite{b29}focused on propagating blocks efficiently. When a new block is generated, in addition to being gossiped across the existing blockchain network, the block is broadcasted to an endpoint in the relay network. The relay network, similar to existing CDNs, broadcasts this block to all other network endpoints where other blockchain network validators can receive the new block if they had  not already seen it. Relay networks usually are able to propagate blocks faster than standard gossip protocols primarily because they directly broadcast new blocks to listening validators. Due to blockchain’s peer-to-peer nature, it is difficult for each validator to efficiently know a large subsection of other validators to directly broadcast information, because of network churn and node discovery. Furthermore, a centralized relay network can implement routing protocols to decrease latency and efficiently route data due to its knowledge of network topography.

Current relay networks include Bitcoin’s Falcon, FIBRE, Bitcoin Relay Network, and the more general Bloxroute network. Each of the aforementioned networks is a relay network, with small efficiency differences between them. One of the primary reasons so few relay networks exist is a result of the network run cost and the fact that many protocols reward validators who produce a block that is not incorporated (unlike Bitcoin). However, this cryptoeconomic design decision is implemented at the expense of efficiency. Of these, only Bloxroute has designed a network that is able to be used by any ecosystem, with ecosystem-owned peer nodes feeding blocks into Bloxroute relay network servers and peer nodes validating blocks as they leave the relay network. Peer nodes ensure the Bloxroute network does not tamper with block data by feeding and verifying test blocks.

Trusted relay networks form the majority of current blockchain relay networks, especially in Bitcoin such as FIBRE or Falcon. Their architecture is simple and similar to existing CDN architecture, while requiring the trust of their users that the network is not manipulating or withholding blocks. Trustless relay networks, such as Bloxroute, are defined as a relay network where blocks that are being propagated may be corrupted or withheld.

Relay networks are especially useful for automotive applications since they, knowing network topography and conditions, can broadcast blocks efficiently to and off vehicle validators alike. This fact allows blocks to be propagated through adverse edge network conditions via dedicated software defined and hardware paths. Similarly, it allows geographically distant validators to potentially allow ecosystem users to take advantage of existing closed intranets, physical lines, and software defined networks. In essence, relay networks directly tackle high latency and latency variation, both contributing factors to orphaned blocks and forks in an automotive blockchain ecosystem.
\cite{b30}, and the more general Bloxroute network. Each of these networks is a relay network, with small efficiency differences between them. One of the primary reasons so few relay networks exist is due to the cost of running them and the fact that many protocols reward validators who produce a block that is not incorporated (unlike Bitcoin). However, this cryptoeconomic design decision is at the expense of efficiency. Of these, only Bloxroute has designed a network that is able to be used by any ecosystem with ecosystem-owned peer nodes feeding blocks into Bloxroute relay network servers and peer nodes validating blocks as they leave the relay network. Peer nodes ensure the Bloxroute network does not tamper with block data by feeding and verifying test blocks.

\subsection{Cut Through Routing}
Cut through routing, such as seen in Bloxroute and the Falcon Network, is a networking optimization that begins retransmitting a block as soon as the first frames of information have arrived. This approach is especially effective at reducing latency due to increasing blocksizes seen in blockchain networks, such as one megabyte for Bitcoin. By transmitting block data once it arrives and before the entire block is received, each node-to-node link’s latency is reduced. This latency loss is especially evident for nodes that are the furthest from where the original block was created. For automotive use-cases, such a technique can be directly embedded into any protocol, reducing the time needed for large blocks to propagate through vehicles and other players.
\subsection{Unreliable Transport With Correction Codes}
Nodes with high latency can partition networks and lead to an increase of forks. The use of reliable transport protocols like TCP can potentially adversely affect latency, due to the acknowledging or ACKing data that has been reliably transferred. Some protocols like FIBRE use unreliable transport protocols like UDP in conjunction with error correction codes to transmit large blocks through networks, in an effort to optimize for latency decreases. This protocol is popular with high bandwidth but high latency links, such as those across continents, in which FIBRE is heavily employed. Furthermore, this technique is being applied through Google’s QUIC protocol \cite{b65} in libp2p \cite{b64} which powers protocols like IPFS \cite{b66} as well as Ethereum 2.0 (Serenity) \cite{b34}. 

\subsection{Transaction Caching}
One major inefficiency with propagating blocks is that validator nodes may have existing transactions that were incorporated into the block being broadcasted. As such, as those validators receive the new block, transaction data may be repeated leading to more information than necessary being sent. BIP 152 (Compact Blocks) \cite{b69} is a released patch where validators can receive a block template containing transaction identifying information and incorporating local transactions into the block. Any transaction not locally present is then downloaded from the validator node which gave the block template. As such, only required information is downloaded between nodes, leading to less information being propagated through the network.

\section{Sharding}
\subsection{Introduction}


Sharding is a blockchain concept where each validator only processes a subset of transactions and/or stores a subset of the global chain. Nodes that maintain a shard maintain information only on that shard while some mechanism exists for handing intershard transactions, which is an active research problem. This phenomenon allows different subsets of validator nodes to split different portions of the global state using the same amount of resources. However, sharding often comes at the cost of compromising trust, usually with the objective of corrupting a single shard. For example, a non-sharded chain with a number of validators agrees to hard-fork into a sharded chain, and splits that number of validators across ten shards, each shard now only has that number of validators divided by the number of shards. This means corrupting one shard only requires corrupting $5.1\%$ or $\frac{51\%}{10}$ of the total number of validators, or usually $ 33\%$ if practical byzantine fault tolerance style consensus (PBFT) based. However, this is usually only dangerous if the validator distribution mechanism is not random. Almost all sharding designs today rely on some source of randomness to assign validators to shards. Note that sharding is usually a protocol layer scalability enhancement even though its goal is to improve network efficiency and locality proximity. Sharding is a major architecture detail for many protocols such as Near Protocol and Serenity. Sharding plays a critical role in increasing throughput for Ethereum’s slow throughput, as seen in Serenity.

Sharding is typically characterized by its quadratic beSharding is typically characterized by its quadratic behavior, that is an increase of throughput in each shard multiplicatively increases the total throughput. Sharding, in its simplest sense, is characterized by Vitalik Bulterin’s ”scaling by a thousand alt-coins”\cite{b33} where small chains of various ecosystems all provide services for their specific users, this approach is taken by Cosmos\cite{b70} and their Cosmos hub as well as Polkadot\cite{b71}. The primarily challenge of sharding is the ability to integrate cross-shard transactions to give the appearance of one large system. Finally, sharding protocols are usually characterized by beacon chains \cite{b32} or inter-shard committees which maintain and govern the shards. Beacon chains also can significantly affect performance due to their inter-shard nature, however they play a critical role with validator assignment, randomness generation, and other inter-shard governance requirements. A temporary epoch based version of a beacon chain is called a committee or inter-shard committee, where normal validators may be chosen to join this committee.


\subsection{Sharding Architecture and Inter-Shard Consensus}
 All sharding protocols are designed to increase transaction throughput rates by dividing the ecosystem into shards that handle transactions within each shard followed by usually some global consensus to achieve state consistency. With automotive and mobility being naturally geographically distributed, the need for sharding becomes apparent. We illustrate this scalability method by example through Elastico, a rotating shard based sharding protocol. As the number of validators in Elastico increases, its transaction rates are expected to increase as well. Specifically, Elastico’s design allows its transaction rates to approximately double with every few hundred nodes added to its network \cite{b40}.
Transaction sharding must address four main laterals:
\begin{enumerate}
\item Validator identification and shard formation
\item Shard leader discovery 
\item Intra-shard consensus
\item Inter-shard consensus and final block broadcast
\end{enumerate}

Upon joining, a network of potential validators must be assigned to a shard. Elastico does this by having potential validators solve a proof of work style verification puzzle to prevent Sybil attacks. After each validator is assigned a shard based on the last, the protocol begins an epoch, or set interval of time, for each shard to exist. Within each shard, Elastico \cite{b40} processes each transaction by using PBFT\cite{b44}. It is important to note that any form of consensus can be used. After the epoch finishes, each shard synchronizes global state over another round of PBFT. Elastico introduces randomness by assigning each validator node randomly to a new shard and repeating the process over. Note that there are many other consensus mechanisms that can be used at this stage. Near Protocol \cite{b35} uses a directed acyclic graph (DAG) to link intershard transactions. This randomness is a crucial factor in preventing byzantine fault for PBFT consensus based protocols, where a fewer number of validator nodes within each shard collude to unfairly manipulate transactions.

Example protocols in this dual PBFT style are Near Protocol, Zilliqa \cite{b36}, RapidChain \cite{b37}, Elastico  which first introduced the overall approach, and OmniLedger \cite{b38}. OmniLedger is a general efficiency and security improvement over Elastico, with the caveat of not having validator reassignment after an epoch. OmniLedger justified this by having a large enough shard, that "one-third" attack become difficult. RapidChain is a general efficiency improvement and offers a decentralized bootstrapping method for initial network genesis. Next, Ziliqa uses a novel signature generation scheme over the other protocols. Finally, Near Protocol uses a directed acyclic graph (DAG) to mitigate orphan block problems between shards due to latency and other network factors, allowing it to mint new blocks faster. Prysmatic Labs uses a similar approach to Near, but implementing their work as a separate protocol based on Serenity \cite{b34}.

\subsection{Static Sharding}
Static sharding randomly assigns validators to shards and does not reassign them for the duration the validator in question is active. Validators may leave and rejoin a network to go through the assignment process again. One major benefit to this approach is the additional latency faced at the end of an epoch due to shard reassignment is automatically mitigated allowing for continuous network activity. A major flaw with this approach is the fact that malicious validators have a smaller pool of validators to corrupt. A malicious entity -- should they be able to predict shard assignments -- would know where potentially hostile validators would be assigned. For many PBFT or BFT based protocols, which generally have one third fault tolerance, this fact lowers the target shard's minimum corrupt faulty validators  to a third of the number of validators divided by the number of shards. As such, strong randomization and fewer shards are necessary. Fewer shards ensure that more validators are assigned to each shard resulting in a malicious actor needing to corrupt more validators. OmniLedger uses initial random validator assignment and increases the number of validators for each shard. This architecture choice allows for more resilient shards with lower failure rates. Additionally, network enhancement protocols can be paired with static sharding to route messages to validator nodes more efficiently. Near Protocol also uses a form of static sharding. Static sharding is especially useful for automotive and mobility usecases, such as the identity use-case, due to the reliance on fewer but larger shards. As such, static sharding has poor performance where data intensive throughput or high transactions are needed. 

\subsection{Epoch Shard Rotation}
Most sharding protocols feature epoch based rotated validator sets. After a predefined epoch ends, the beacon chain reassigns validators to new shards and then transactions resume. One major benefit of this property is that predicting which shard a validator will be assigned to becomes as difficult as predicting the randomness protocol that governs assignment. However, this feature is at the expense of latency with the entire system not being available to process new transactions during validator assignment. For continuous or real time applications that require constant system availability, this system would not naively be applicable. However, for any other applications which tolerate predicted though not necessarily bounded delays, such a system is applicable. Examples of epoch based shard rotation protocols include Elastico, Zilliqa, Ethereum Serenity, and RapidChain. One major problem with epoch based shard rotations is the associated system freeze time when an epoch ends. However pipelining random values sufficientily solves this problem, as seen in Ethereum Serenity's RANDAO \cite{b68}.
\subsection{Randomness in Sharding}
One of the major responsibilities assigned to beacon chains or inter-shard committees is randomness generation. Randomness makes guessing "behavior" more difficult, probabilistically reducing many attacks such as the $\frac{1}{3}$ fault tolerance attack in PBFT protocols and 51\% attack in proof of work consensus style protocols. A variety of randomness inducing methods exist specifically for sharding. OmniLedger, a major static sharding based protocol, uses RandHound\cite{b42}, a multiparty computation protocol (MPC), to deliver byzantine fault tolerant randomness. Elastico uses a distributed commit and xor scheme. Other protocols use the output of previous steps in randomness generation. Zilliqa uses the output of a proof of work nonce to assign validators to shards. Serenity uses verifiable delay function based randomness within their beacon chain through a decentralized autonomous organization (RANDAO)\cite{b68}. Serenity additionally pipelines backup RANDAO values to prevent system stall as discussed in the previous subsection. 

\subsection{State Sharding}

\subsubsection{State Splitting}
Sharding transactions is not enough to ensure data efficiency, due to the need for state to be stored such that it is consistent throughout the entire sharded system. Some protocols, like Zilliqa, ignore state sharding completely, at the benefit of not needing to implement inter-shard transactions. However, most protocols implement state sharding such that most validator nodes only store transactions that affect transactions processed by that shard. Most non-polynomial encoded state sharding protocols do not share state among all shards, meaning a ledger of all transactions is not stored at each validator. However, network consistency must be able to be proved at the inter-shard level. Elastico stores transaction data for only transactions that occurred within the shard. However, the Merkle roots from each shard are merged by way of a cryptographic hash function by the inter-shard committee. This allows a validator to verify state at any time by requesting each shard's Merkle root from that epoch and re-run the hash function to recreate the inter-shard's committee's value for that epoch. The primary benefit for such an approach is that each shard only keeps data relevant to transactions within or affecting accounts or other state data within a shard. The vast majority of protocols follow this template with where only relevant shard data. 
\subsubsection{Inter-Shard Consensus}
One major challenge with split state shards is resolving transactions that affect more than one shard, due to the fact that transactions affect two distinct processing divisions. Elastico, like most other sharding protocols asynchronously executes inter-shard transactions with each transaction partially occurring within relevant shards. There are two main approaches to this problem. The less used synchronous method where cross shard transactions are simultaneously coordinated and executed by both blocks of the two shards in question. The asynchronous method where the two shards asynchronously execute transactions but rely on one shard demonstrating a transaction executed for the other shard to execute the transaction, with this approach being primarily used for its flexibility and lack of forcing the two shards to synchronize in parallel , preserving performance.  The train and hotel problem specified by Andrew Miller \cite{b33} highlights the general template of asynchronous cross-shard transactions, where transactions are modelled through dependency graphs with shards executing dependencies first. Serenity, Elastico, and most protocols use this approach. One major future direction is the incorporation of ZK-STARKS and ZK-SNARKS\cite{b33} to create receipts of dependencies executed and thereby allowing the next dependency to proceed. However, most current ZK-STARK and ZK-SNARK implementations are slow resulting in  excessive latency leading to inter-shard consensus being an area of active research. Especially for mobility, being able to tolerate inter-shard transactions with reasonably high throughput is important where user shard assignment is executed randomly. 

\subsubsection{Polynomial Erasure Data Coding}
Polynomial encoding and erasure codes permit data to be recovered given enough points for reconstruction. Such an approach can be used for efficiently encoding past transaction data and creating enough data shards such that data reconstruction can be done given node failure.  We use Polyshard \cite{b40} to illustrate polynomial encoding based state sharding, an efficient method for state sharding. Note that as of now, Polyshard and erasure coding from Serenity are the only significant public works in the field. Polyshard mitigates the previously described issue with each node storing and computing on a coded shard of the same size that is generated by linearly mixing uncoded shards, using the  Lagrange polynomial function. This function provides computational redundancy as well as provides security against erroneous results from malicious nodes, enabled by noisy polynomial interpolation techniques, through Reed-Solomon encoding. This sort of encoding is similarly extended to Serenity's related data availability usecase, which will incorporate erasure codes for data availability for fraud proofs\cite{b41}; this use exemplifies the compactness of polynomial encoding allowing even light clients greater capacity to participate. 

While polynomial encoding is generally applicable in many distributed computing scenarios, the following two salient features make PolyShard and by polynomial encoded  state sharding particularly suitable for blockchain systems in mobility:
\begin{enumerate}
\item Oblivious: The coding strategy applied to generate coded shards is independent of the verification function. That implies that the same encoded data can be simultaneously used for multiple verification items (for example: digital signature verification and balance verification in a payment system)

\item Incremental: PolyShard allows each node to grow its local coded shard by coding over the newest verified blocks, without needing to access the previous ones. This helps maintain a constant coding overhead as the chain grows, allowing scalable and secure solutions to be utilized. 
\end{enumerate}


State sharding is critical factor to scaling storage. Bitcoin, one of the few large-scale blockchain systems, currently requires a full 185 gigabytes of current blocks to be stored on each full validator node. In a mobile environment with theoretically one billion vehicular participants, state sharding, similar to Polyshard, ensures that every validator does not have to contain the full set of all blocks and capacity grows with each node joining the network. This allows validation to be decoupled, due to storage constraints, from large data centers to other potential actors within a mobile ecosystem, such as vehicles themselves. It can be a powerful tool to allow other mobile ecosystem players to have control or simply reduce storage costs.
\subsection{Automotive Sharding Analysis}
\subsubsection{Data Efficiency}
Sharding transactions does not natively solve the space constraint issue. Since the entire network must be able to reach consensus, as dictated by network safety, each validator will usually contain the entire networks history. Approaches such as state sharding can be used to solve this issue. For automotive architecture, this is critical to storage management since most vehicles do not have the capacity to have large datacenter like storage capabilities. Even if validators are constrained to non-vehicles, failing to scale data storage with shard division as in Zilliqa's approach would result in unnecessary data replication and would result in a highly inefficient network. 

\subsubsection{Shard Latency}
Additionally, intra-shard consensus may slow down confirmation times for transactions that took place between two shards, as seen in the Near protocol. This delay is unavoidable, though possibly mitigated by grouping participants who commonly exchange commit transactions with each-other together in the same shard. The key to automotive is the exchange between validator assignment randomness and latency. For example, a validator may be physically far away such that the latency delay between a network participant and validator may become too high, especially for mission critical tasks. Potential approaches include creating smaller shards and pseudo-randomness such that validators are always within an exact or estimated latency bound. However, such a scheme or similar schemes allow potential validators to narrow, sometimes significantly the set of network participants they will be serving, introducing potential crypto-economic risks. Many other approaches such as using a class of semi-trusted verifiers or various staking mechanisms may be appropriate to mitigate risks.

\section{Consensus Models}
Consensus models form the heart of any automotive blockchain system, providing the final agreement of global state. While consensus has been a major distributed systems topic for years, Nakamoto style consensus \cite{b16} (longest chain consensus) was the first major blockchain focused consensus protocol. However, its inefficiency due to its reliance on proof of work \cite{b43} led to a resurgent focus on practical byzantine fault tolerant consensus (PBFT) \cite{b44}. BFT's key difference is prioritizing consistency of transactions over system availability as in Nakamoto style systems. Tendermint\cite{b18} and Algorand \cite{b19} were among the first widely known PBFT protocols introduced. However, PBFT algorithms, due to the "voting" structure of this class of consensus algorithms, are more susceptible to issues of liveness due to network issues. In automotive use-cases, the commonality of arbitrary network delays, expensive data transfer costs, temporal connections, as well as mobile ecosystem players introduce difficulties for many current consensus mechanisms used to stable static validators and ecosystem agents.

\subsection{Directed Acyclic Graphs}
Directed acyclic graphs (DAGs) present a relevant solution to directly tackling the orphan block problem \cite{b49}, a major issue for automotive due to variable latencies associated with validation on-board vehicles. DAGs do not directly influence a systems scalability, instead DAGs are a data structure replacement to the chain data structure seen in most other protocols. By allowing orphan blocks that are incorporated into the DAG, direct scalability methods such as increasing block size, allowing more collections to be processed for the block period, or simply increasing the block creation rate. Note that the vast majority of DAG based protocols still require partial synchronicity, guaranteeing that messages are eventually delivered through the network even if the upper bound delay is not directly known. This particular factor makes DAGs especially suitable for mobility and automotive based use-cases, where higher changes of network partitioning and delay are likely to happen. One major drawback to DAGs is the difficulty in calculating the probability of transaction reversal due to non-linearity. Additionally DAGs are prone to network inconsistency due to blocks not being propagated network wide before a conflicting transaction may occur. 


\subsubsection{Heaviest Path}
The most similar DAG analogue protocol to Bitcoin's blockchain using Nakamoto style consensus is the Ghost protocol \cite{b49}. Similar to Bitcoin style blockchain, where the longest observed chain is mined on by honest validators, Ghost encourages miners, using Nakamoto style proof of work mining, to append blocks onto the heaviest observed subtree in the DAG. Each time a validator receives the entire DAG, the validator traverses the entire DAG choosing the path with the heaviest weight. In Nakamoto style consensus, this translates to the subtree with the most nodes linked. Essentially, this is the subtree where the most mining was previously completed. Safety is achieved based on the same principle of Bitcoin's blockchain, being that in the time it takes for the newly minted block to propagate through the entire network the validator will either have received a new valid block or every node in the network will have received the original minted block. 

\subsubsection{Vote Ordering}
Building on Ghost is the Spectre protocol \cite{b50}. Spectre's primary novel contribution is the ability to find a ground truth in transactions by relying on transactional precedence of a DAG's voted on topological sort. The primary benefit of DAGs and chains is that both can be reduced to a distinct ordering of transactions, which in turn ensures that no new transaction can conflict with a previous transaction. Spectre as a consensus mechanism has each validator vote on the topological sort of the DAG, therefore ensuring a consistent state. Transactions are still processed and added onto the DAG the same way Ghost protocol does so. The decoupling of mining and transaction consensus allows Spectre to process far more transactions on the network. One of the primary issues with Spectre is that the voting protocol does not inherently ensure that the topological sort is valid, since the ordering itself is voted on rather than the DAGs structure. This issue more practically manifests itself with the fact that transactions cannot be strongly ordered, resulting in Ethereum Virtual Machine (EVM) style transactions not being possible. As such, this platform is not ideal for automotive use-cases, where many complex transactions are not commutative.  However, in some use-cases where transactions do not require strict order, Spectre provides an exceptionally simple and fast DAG paradigm. By switching the proof of work mechanism with a more efficient stake based system, a reasonable and simple mobility focused protocol could be built. 

\subsubsection{Node Clustering}
Phantom \cite{b51} builds on Spectre by removing voting based consensus in favor of calculating the correct path from leaf to root of blocks. This method, while more computationally expensive, ensures that the DAG's acyclic integrity is always maintained and enforces a strict ordering of transactions, thereby allowing complex non-commutative transactions. State correctness is achieved by calculating the maximum k sized cluster sub-DAG. This sub-DAG colloquially is the set of blocks where for any block the number of paths to and from that said block is bounded by k. In essence, it prioritizes the portion of the DAG that is more connected and therefore less susceptible to attacks like chain holding. The newly minted block then connects all leaves in the k cluster sub-DAG further penalizing chain withholding validators. Note that the k clustering problem is NP-hard and therefore an approximation is used to estimate this set of blocks. As such, transaction precedence and correctness can then be calculated by topologically sorting the "correct" portion of the DAG and checking for transaction conflicts, rejecting new transactions that conflict. This DAG method is particularly suitable for complex transactions involving automotive use-cases due to the ability to include non-commutative transactions. As with Spectre, the protocol in the paper uses proof of work, however this may be switched with other stake based systems if adapted for use in larger automotive systems. One major drawback with Phantom is that it explicitly penalizes validators with high network latency due to their ability to be less "connected" then validators with lower latencies who are able to reference more blocks in time. The Inclusive protocol \cite{b52} from the same authors introduces a novel mechanism called the "Inclusive Game" which relies on fees to incentivize transactions to be rationally non-conflicting while still allowing more transactions to be included in the DAG. This replaces the k cluster algorithm in Phantom. 

\subsubsection{Federated Byzantine Agreement DAGs}
Another novel variation of using DAGs in consensus protocols is the Avalanche family \cite{b53} (Avalanche, Snowball, Snowflake, and Slush). Similar to Spectre, Avalanche utilizes vote based ground truth which is sampled from other validators. The Snowball portion of the paper introduces the idea of calculating confidence in some state, visualized as a coloring of a node, to be one of two colors. The network is randomly sampled until confidence is passively voted on based on another node's states. This fact additionally establishes leaderless passive consensus, negating the need for complex vote based rounds and message passing which can be risky liveness-wise for many automotive based blockchain use-cases, due to the reliance on a round leader. Avalanche builds on snowflake by storing this confidence of a transaction by storing a transaction's confidence in a block in a DAG structure. The strength of a transaction is determined by validating all previous transactions that are reachable through the DAG structure. This can be seen as the same as ensuring that all previous transactions that led to the current state are correct therefore this transaction is correct by not conflicting a previous transaction. Block generation is done by pooling a set of transactions waiting to be confirmed into the same block then running the correctness validation protocol. One of Avalanches greatest strengths is its efficiency, achieving up to 1300 tx/s compared to 364 tx/s with Algorand when tested in a single EC2 cluster with 2000 validators and negligible network latency. Additionally, by negating the need for consistent proof of state or proof of work (a benefit of leaderless protocols), Avalanche achieves $10^{-9}$ probability of an adversarial transaction being accepted, similar to proof of stake based Algorand's $5 x 10^{-9}$ with $20\%$ of Byzantine nodes \cite{b19}. When paired with a strong validator identity mechanism, though not required, Avalanche makes a strong automotive and mobility network case, where honest transactions are guaranteed commitment in conditions similar to other PBFT style protocols and efficiency due to the lack of formalized round based voting. One current issue with Avalanche is that the paper does not specify how parent transactions should be chosen, resulting in potentially controversial transactions never being voted on and approved. This issue is a result of new transactions being built on validated old transactions. A sufficient parent selection algorithm would mitigate this flaw.

In summary, DAGs due to their tolerance for network delays  leading to orphaned blocks can form a strong latency tolerant part of any protocol for automotive use cases. 
\subsection{Optimistic Path}
Optimistic Path style consensus achieves consensus by combining a fast, asynchronous path with instant transaction confirmations with a fully synchronous “fall-back” path, which the system falls back to if transactions are disputed. This architecture provides a simple paradigm that in adversarial conditions is as robust as most synchronous protocols \cite{b54}, yet when given a super majority of honest validators optimistically confirms transactions. Optimistic path consensus remains resilient assuming only that a majority of the computing power is controlled by honest players. Optimistically when $\frac{3}{4}$ of the computing power is controlled by honest players,  a special player called the “accelerator” is used to confirm and append transactions instantaneously, bounded only by the message delay in the network. Additionally, it can be shown that the $\frac{3}{4}$ optimistic bound is tight for protocols that are resilient assuming only an honest majority. Thunderella \cite{b54} is an example of a protocol that employs optimistic path consensus. Optimistic path consensus is especially beneficial where an ecosystem has a heuristic determining that participants usually will act honestly. However, an overly hostile or cheating ecosystem will be greatly inefficient under this system. For systems where external knowledge is known about validators and participants, such as automotive, and where a strong heuristic can be applied, optimistic path consensus remains an efficient option which effectively utilizes external environment information in its heuristic. 

\subsection{Hardware Trusted Execution Environments }
Hardware based trusted execution environment (TEE) and their associated protocols represent a distinct set of benefits and potential drawbacks for automotive use-cases. Hardware based TEEs, such as Intel Corporation's SGX (software guard extension) launched with 2015 with their sixth generation Skylake family \cite{b14}, offer aTEE implemented in hardware as opposed to an OS's kernel or application running on an OS. SGX offers hardware isolated memory for applications that utilize them, even protecting against a malicious kernel. This practically allows the storage of secrets in memory. An interface abstraction allows the use of special machine code commands to marshal and un-marshal data between the enclave and un-trusted program. However, any single flaw in the enclave renders the entire system compromised, with attacks such as the spectre attack \cite{b55} famously exploiting an efficiency mechanism in the CPU while using unprivileged access (non-kernel access) to dump enclave contents into unprotected memory. Flaws in SGX are difficult to fix if they cannot be patched using firmware due to the necessity of replacing the full CPU. However for automotive use-cases where tier one suppliers and automakers completely control hardware and its servicing, such mistakes may be easier to replace and manage. For this reason, hardware based TEEs like SGX make a compelling case for automotive adoption. 

Exploiting SGX's memory isolation is Intel's novel Proof of Elapsed Time (PoET) protocol \cite{b57}. It extends traditional Nakamoto style consensus, mimicking proof of work with a timer run in an SGX TEE. Rather than continuously hashing values to solve the cryptographic puzzle traditionally seen in Nakamoto consensus, POET has validators generate a random number representing the time before the current block can be mined. The validator with the lowest wait time mines and broadcasts the block before the other validators with the process then restarting. One major flaw with this protocol is that it requires every validator to prove their identity and that they have the appropriate SGX settings enabled. As such, this consensus protocol is only possible on ecosystems where identity is guaranteed with a strong probability. For automotive applications this guarantee may be acceptable due to the strict control of vehicle identity and hardware, however it poses a limiting factor where identity control is not possible. POET is implemented in Hyperledger Sawtooth, which in turn is part of the greater Hyperledger ecosystem. 

Ekiden\cite{b56} builds on hardware based TEEs by eschewing POET/Nakamoto consensus completely. Instead Ekiden decouples code execution with consensus, running logical code inside the SGX TEE which in turn passes a proof of publication. A process flow for Ekiden has a smart contract \cite{b56} submitted to an execution node which contains the TEE which publishes the contract to a consensus node which appends the contract onto the blockchain. Should a smart contract be activated, the compute node again processes the smart contract information and then publishes any resultant state changes to the consensus nodes which in turn update the blockchain. 
\subsection{Round Pipelining}
Round pipelining refers to having two different round's data stored on the same message broadcasted to the entire set of validator nodes. This simple pipelining mechanism is succinctly discussed in detail in PaLa \cite{b58}, a simple PBFT protocol that implements pipelining. Pipelining is especially useful as a simple efficiency boost, due to the ability to run more than one round concurrently over the same network messaged passed to and from the leader of a round. As of this paper's publication, there is no evaluation of protocol. An implemented protocol which uses round pipelining is Dora \cite{b67}, one that extends Tendermints's PBFT consensus protocol \cite{b18} Tendermint with pipelining. While the overarching concept remains the same, the specifics of implementation for each protocol remain unique to the underlying consensus mechanism as seen in Dora and PaLa. 

\section{State Channels}
Application layer scalability protocols such as state channels and Ethereum's Plasma \cite{b25} are unique from the other methods discussed in this paper as they exist above any network or protocol level enhancement. As such, they are platform agnostic and far more generalizable than the other approaches discussed. However this generalizability comes at a technical trade-off for usability and complexity. State channels are particularly useful for deposits used over time without withdrawal. However, many use cases such as automotive payments or products involving a temporal recurring fee make attractive applications of state channels. 

State channels at their most basic form allow two users to lock tokens or other stateful components, as in Counterfactual \cite{b72}, into a smart contract and maintain an off chain ledger between the two users of active transactions only between those two users. Once the users deem an epoch finished, they both sign a final "state" of the ledger and submit this ledger as a final transaction signed by both parties to the smart contract which disperses tokens or components on-chain. This simple paradigm forms the base of all state channel networks such as the Lightning Network \cite{b23} for Bitcoin and Raiden Network \cite{b24} for Ethereum. The ledger kept between two participants can also generalized to any multi-agent object, with the smart contract needing to be able to transform this object into a specific payout between two participants \cite{b60}. This paradigm also highlights one of the greatest drawbacks of state channel based systems, state channels are only efficient if there exists a significant number of transactions between two parties. Should both parties constantly close an active contract and form new contracts, the system devolves into a large set of on-chain transactions removing the highly scalable off chain advantage. Both Lightning and Raiden attempt to mitigate this architecture problem by constructing more complex state channel networks such as multi-hop and hub and spoke channel systems \cite{b23} \cite{b59}. Both approaches rely on a network of participants who have active contracts with each other. When a transaction needs to be made from participant A to B, a path of nodes from A to B such that each node has enough tokens staked for each adjacent node in that path. With this path B generates a secret and its hash and passes the hash to A. A then gives a payment promise and the hash the first node in the path which is then propagated all the way to B, with B then passing the original secret the other way up the chain allowing the path to fulfill payments. 

One of the most apparent difficulties with such a system is ensuring connectivity between participants. Hub and spoke architecture was developed \cite{b59} to minimize the length of paths and ensure connectivity by using a few fully connected hubs to all users such as seen in Finality Labs \cite{b59}, however the overall process flow involves significant potential latency related to messaging and path discovery. Furthermore, state channel directional architecture plays the greatest role in efficiency. Participants in the hub and spoke model must initially be willing to stake tokens and stateful components and hold them in that contract for a period of time. For automotive usecases, this is of particular interest involving payment usecases where money is initially deposited and used over time without withdrawal. In particular, multiple hubs offer a way for transactions over a defined epoch to offer the same decentralized and cryptographic guarantees as normal blockchain transactions directly built on a protocol without the need for consensus until an epoch is over. For this reason, state channels offer the most attractive path for payment systems should hub and spoke architecture be implemented correctly with high enough liquidity. 

\section{Scalability Use-Case Applications}
We now explore applications of the discussed scalability techniques. Note that section is a small and not exhaustive list, with many other use-cases existing. 
\subsection{Payments}
The peer-to-peer payments use case  involves any sort of payment to and from vehicles to other vehicles and infrastructure. Due to the simplicity and need for almost real time confirmations, hub based state channel architecture allows rapid payments between participants without the need to constantly update the underlying blockchain state. Without explicitly forcing consensus except at the end of an epoch a high degree of efficiency is achieved with only nodes along the payment path explicitly being involved. 
\subsection{Ownership and Identity Management}
Ownership tracking and identity management are both proposed solutions toward creating a fully digital and decentralized method for tracking ownership and identity, replacing the patchwork of systems currently used. One important caveat is that state consistency is important, with slow propagated state updates resulting in incorrect permissions. Such a system would greatly benefit from the  routing enhancements discussed in Section III, enabling light clients to quickly gain access to updated information and therefore maintain state. With faster state propagation, ownership and identity solution can now interact with ecosystem players near real time, depending on implementation. 
\subsection{Black Box Data Sharing and Processing}
Black box data sharing and processing refers to a general class of applications that allow different entities to share data or process data from another organization. Suggested automotive applications involve granular driving habit analysis for dynamic insurance premiums as well as distributed autonomous vehicle training. Ekiden's SGX based compute nodes execute smart contract code on the particular node in question then prove correct execution on chain. One of the biggest benefits for this model is efficiency, allowing complex algorithms to be run on local data without requiring every compute node to run the same instructions. Furthermore, data would never physically be transferred to a requestor's systems, allowing for more private systems and potentially making regulatory compliance more simple. As such TEE hardware provides a clear improvement with regards to scalability and privacy. 

\section{Conclusion}
Blockchain scalability is one of the greatest hurdles preventing its usage in the modern automotive and, more generally, the mobility sector. However, a balance or compromise according to the application specific requirements can be made in order to exploit blockchain in specific circumstances. Scaling blockchain systems that require sending and processing large amounts of data should allow some degree of compromised security and centralization with a blockchain based protocol can be designed accordingly. Similarly if user necessities are different such that increased adversarial conditions is needed, the protocols can similarly be changed. For example, Rippless network may be able to achieve better scalability, but it comes with increased centralization and additional issues like rotated trusted validators. Bitcoin Cash, on the other hand, boosted scalability by raising blocksize limits, but the higher blocksize limits may compromise the security of the network. Ensuring scalability with appropriately chosen security modeling is the only way for blockchain technology to be accepted for automotive use cases.



\vspace{12pt}

\end{document}